%Paper: astro-ph/9403044
%From: bss@iucaa.ernet.in (Sathyaprakash)
%Date: Mon, 21 Mar 94 16:57:47+050

\magnification 1200
\hoffset .5 true in
\hsize = 6.0  true in
\vsize = 8.0 true in
\parskip = 6pt			% leaves space after para
\font\tentworm=cmr10 scaled \magstep2
\font\tentwobf=cmbx10 scaled \magstep2
\font\tenonerm=cmr10 scaled \magstep1
\font\tenonebf=cmbx10 scaled \magstep1
\font\eightrm=cmr8
\font\eightit=cmti8
\font\eightbf=cmbx8
\font\eightsl=cmsl8
\font\sevensy=cmsy7
\font\sevenm=cmmi7
\font\twelverm=cmr12
\font\twelvebf=cmbx12

\def\subsection #1\par{\noindent {\bf #1} \noindent \rm}
\def\mid {\let\rm=\tenonerm \let\bf=\tenonebf \rm \bf}
\def\para{\par \vskip 12 pt}
\def\head{\let\rm=\tentworm \let\bf=\tentwobf \rm \bf}
\def\heading #1 #2\par{\centerline {\head #1} \smallskip
\centerline {\head #2} \vskip ./5 pt \rm}
\def\eight{\let\rm=\eightrm \let\it=\eightit \let\bf=\eightbf
\let\sl=\eightsl \let\sy=\sevensy \let\m=\sevenm \rm}

\def\foots{\noindent \eight \baselineskip=10 true pt \noindent \rm}
\def\sexion{\let\rm=\twelverm \let\bf=\twelvebf \rm \bf}

\def\section #1 #2\par{\vskip 20 pt \noindent {\mid #1} \enspace {\mid #2}
\para \noindent \rm}

\def\ssection #1 #2\par{\noindent {\mid #1} \enspace {\mid #2}
\para \noindent \rm}

\def\abstract#1\par{\para \foots {\bf Abstract: \enspace}#1 \para}

\def\author#1\par{\centerline {#1} \vskip 0.1 true in \rm}

\def\abstract#1\par{\noindent {\bf Abstract: }#1 \vskip 0.5 true in \rm}

\def\midsection #1\par{\noindent {\sexion #1} \noindent \rm}

\def\sqr#1#2{{\vcenter{\vbox{\hrule height#2pt
\hbox {\vrule width#2pt height#1pt \kern#1pt
\vrule width#2pt}
\hrule height#2pt}}}}

% definitions
\def \dk {\vert\delta_k\vert}
\def \dk2 {{\vert\delta_k\vert}^2}

\def \half {1\over 2}

\def \ap {ApJ, }
\def \apl {ApJ, }
\def \aps {ApJS, }

\def \prl {Phys. Rev. Lett., }
\def \pl {Phys. Lett., }

\def \mnras {MNRAS, }
\def \aa {A\&A, }

\def \doublespace {\baselineskip = 20pt plus 7pt \message {double space}}
\def \singlespace {\baselineskip = 13pt plus 3pt \message {single space}}
\singlespace

\def \body {\vfill \eject \parindent = 1.0 true cm
	\ifx \spacing \singlespace \singlespace \else \doublespace \fi}

\def\head{\let\rm=\tentworm \let\bf=\tentwobf \rm \bf}
\def\heading #1 #2\par{\centerline {\head #1} \smallskip
\centerline {\head #2} \vskip .15 pt \rm}
\def \title#1 {\centerline {{\bf #1}}}
\def \Abstract#1 {\noindent \baselineskip=15pt plus 3pt \parshape=1 40pt310pt
  {\bf Abstract} \ \ #1}

 %large size Roman for titles
 %parallel

\catcode`@=11
\def \C@ncel#1#2 {\ooalign {$\hfil#1 \mkern2mu/ \hfil $\crcr$#1#2$}}
\def \gf#1 {\mathrel {\mathpalette \c@ncel#1}}	% slash a small letter
\def \Gf#1 {\mathrel {\mathpalette \C@ncel#1}}	% slash a big letter

\def \gapx {\;\lower 2pt \hbox {$\buildrel > \over {\scriptstyle {\sim}}$}
\; }
\def \lapx {\;\lower 2pt \hbox {$\buildrel < \over {\scriptstyle {\sim}}$}
\; }
\topskip 0.7 true cm

\def\n{\noindent}

\def\v2{\vskip 0.2 true cm}
\def\v3{\vskip 0.3 true cm}
\def\v4{\vskip 0.4 true cm}
\def\ie{{i.e., }}
\hyphenation {dimen-sional}

\baselineskip=0.3333truein plus 0.05true in % defines line spacing for ApJ

\heading {The Evolution of Voids in the Adhesion Approximation}

\vfill
\vskip 1.0cm
\centerline{\bf{Varun Sahni$^1$, $\,\,\,$ B.S. Sathyaprakash$^1$ }}
\vskip 0.2cm
\centerline{and}
\vskip 0.2cm
\centerline{\bf{Sergei F. Shandarin$^2$}}
\vskip 0.5cm
\centerline{$^1$ Inter-University Centre for Astronomy and Astrophysics}
\centerline{Post Bag 4, Ganeshkhind, Pune 411 007, India}
\vskip .2cm
\centerline{$^2$ Department of Physics and Astronomy}
\centerline{University of Kansas, Lawrence, Kansas 66045, U.S.A.}
\vskip 0.5cm
\centerline{\bf To appear in The Astrophysical Journal July 1994}

\vfill
\bigskip
\topskip 1.0cm
\n
\centerline{ABSTRACT}
\vskip .5cm
We apply the adhesion approximation to study the formation and evolution
of voids in the Universe.
Our simulations -- carried out using 128$^3$ particles in
a cubical box with side 128 Mpc --
indicate that the void spectrum evolves with
time and that the mean void size in the standard COBE-normalised
Cold Dark Matter (hereafter CDM) model
with $h_{50} = 1,$ scales approximately as
$\bar D(z) = {\bar D_0\over \sqrt {1+z}},$ where $\bar D_0 \simeq 10.5$
Mpc.
Interestingly, we find a strong correlation between the sizes of
voids and the value of
the primordial gravitational potential at void centers. This observation
could in principle, pave the way towards reconstructing the form of
the primordial
potential from a knowledge of the observed void spectrum.
Studying the void spectrum at different cosmological epochs,
for spectra with a built in $k$-space cutoff we find that,
the number of voids in a
representative volume evolves with time.
The mean number of voids first increases until a maximum value is reached
(indicating that the formation of cellular structure is complete),
and then begins to decrease as clumps and filaments merge leading to
hierarchical clustering and the subsequent elimination of small voids.
The cosmological epoch characterizing the completion of cellular
structure occurs when the length scale going nonlinear approaches
the mean distance between peaks of the gravitational potential.
A central result of this paper is that voids can be populated by
substructure
such as mini-sheets and filaments, which run through voids. The number
of such mini-pancakes which pass through a given void, can be measured
by
the {\it genus characteristic} of an individual void which is an
indicator of the topology of a given void in initial (Lagrangian) space.
Large voids have on an average a larger genus measure than smaller
voids indicating more substructure within larger voids relative to
smaller ones.
We find that the topology of individual voids
is strongly epoch dependent, with void topologies generally simplifying
with
time. This means that as voids grow older they become progressively
more empty and have less substructure within them.
We evaluate the genus measure both for individual voids as well as
for the entire ensemble of voids predicted by the CDM model.
As a result we find that the topology of voids when taken together with the
void spectrum,
is a very useful statistical indicator of the evolution of the
structure of the Universe on large scales.

\vskip 1.0cm
\vfill\eject
\topskip  0.7cm
\vskip .5cm
\centerline {I.\phantom{.} INTRODUCTION}
\vskip .4cm
 A remarkable feature of the large scale structure of the Universe is the
 presence of voids -- volumes of order $10^3h^{-3}$ Mpc$^3$, virtually
 devoid of the presence of galaxies. The existence of voids was spectacularly
 demonstrated with the discovery of the Bootes void an
 $\sim 10^5h^{-3}$ Mpc$^3$ underdense region, by Kirshner et al. (1981).
Since then successive redshift surveys (de Lapparant et al. 1986,
Vogeley et al. 1991; Slezak et al. 1993), as well as deep pencil beam surveys
(Broadhurst et al. 1990), have confirmed that voids are a salient feature of
the large scale structure of the Universe, and that galaxies
 seem to lie preferentially along sheets and filaments separating voids.

Although it is generally agreed that voids account for most of the volume
of the Universe, there is still considerable
disagreement in the literature as to what is the typical size of a void.
Part of this disagreement has to do with how one chooses to
define voids. If voids are defined as underdense regions then the size of
the void will obviously depend upon the density threshold below which
an underdense region ``becomes'' a void.
Clearly, the mean volume of underdense regions will always be larger than the
mean volume of completely empty regions.
The average size of a void in a sample will, in addition,
be sensitive to the smallest
voids which one recognizes. Clearly, if we recognize empty regions of diameter
$\sim 0.2 $ Mpc to be voids, then the mean void size will be much smaller than
if we were dealing with voids having diameters $\ge 2$ Mpc.

However, even if we restrict ourselves to voids defined as completely empty
regions, we still come across differing points of view as to the
typical void size in the Universe. Thus, Kauffmann and Fairall (1991)
(henceforth KF), who
consider only void diameters larger than $\sim 5h^{-1} $ Mpc
in an uncontrolled
data sample, find from their catalogue
of over 100 voids that the typical ({\it viz,} modal)
size of a completely empty region is $D \sim 10 h^{-1}$ Mpc. The mean void
diameter in the KF sample is $36.5 h^{-1}$ Mpc
(Little 1992).\footnote{$^1$}{Little has evaluated the mean void diameter for
129 voids determined from the merged Southern Redshift Catalogue
and the Catalogue of Radial Velocities of Galaxies and listed by KF
 in their paper.}
The authors also find no evidence of a cutoff in void size
upto a maximum void diameter of $64~h^{-1}$ Mpc.
On the other hand, Vogeley et al. (1991),
who work with a smaller but controlled sample (the CfA survey), find
the maximum size of a completely empty region in their sample to be
$\sim 20 h^{-1}$ Mpc which is considerably smaller than the {\it mean
void size} in the KF sample.
A part of this discrepency may arise from the fact that KF
work with a large but uncontrolled data sample, whereas the CfA survey
although containing fewer galaxies has the advantage of
being magnitude limited.
(KF use galaxy redshifts which are not magnitude limited since they are
compiled from many different sources, thereby giving rise to the
``distinct possibility that certain regions of the sky are undersampled
in comparison to other regions '' and to the related possibility that
the existence of a particular void is due to undersampling rather than
to any physical process (Kauffmann $\&$ Fairall 1991).)
The other reason why the results of
Vogeley et al. (1991)
and those of KF differ, could be due to the
different techniques used to actually measure void sizes.
Thus, Vogeley et al. (1991)
construct a void probability function for the CfA survey which measures the
probability that a randomly selected volume in their survey contain no
galaxies, whereas KF determine  a void spectrum
by filling regions empty of galaxies in their
sample with cubes, the edge length of the largest cube filling a given
empty region providing a measure of the void size.
Although this method leads to a well defined algorithm to measure void sizes
it suffers from the drawback (acknowledged by the authors)
of ignoring the issue of void topology.
(Voids in the KF sample are assumed a-priori,
to possess a bubble like topology.)

Since voids are associated with positive peaks in the primordial
gravitational potential,
(Gurbatov, Saichev $\&$ Shandarin, 1985, 1989; Kofman $\&$ Shandarin, 1988)
they give rise to anisotropies in the Cosmic Microwave Background
Radiation (CMBR) via the Sachs-Wolfe effect: $(\delta T/ T) \simeq
\delta \phi/3$ on the surface of last scattering.
This effect has been studied for individual voids
by  Blumenthal et al. (1992)
who find the sizes of the largest voids to be smaller
than $\sim 130 h^{-1}$ Mpc (one void of this size within the horizon).
In a later paper the same authors incorporate the COBE results
to give a stronger limit $\sim 60 h^{-1}$ to the maximal void size
(Piran et al. 1993).

 From the theoretical viewpoint
the fact that voids might provide a key to understanding the large scale
structure of the Universe was emphasised more than a decade ago,
by Zeldovich $\&$ Shandarin (1982) and Zeldovich, Einasto \& Shandarin (1982).
Since then, the evolution of voids has been examined both analytically as well
as numerically by a number of authors (Peebles 1982; Hausman, Olson $\&$ Roth
1983; Hoffman, Salpeter $\&$ Wasserman 1983; Fillmore $\&$ Goldreich 1984;
Bertschinger 1985; Blaes, Villumsen $\&$ Goldreich 1990;
Bonnor $\&$ Chamorro 1990; Regos $\&$ Geller 1991; Dubinski et al. 1992;
van de Weygaert $\&$ van Kampen 1993; Harrington, Melott $\&$ Shandarin 1993).
Semi-analytical spherically symmetric studies of voids have shown that large
voids can often arise out of small initial negative density fluctuations.
Numerical simulations by Bertschinger 1985 and
Blaes, Villumsen $\&$ Goldreich 1990, confirmed that aspherical
negative density
perturbations grow to become more spherical with time (Icke, 1984)
 (the time-reverse of the
Lin, Mestel $\&$ Shu (1965) instability), indicating that the evolution of
negative density perturbations can often be treated as the time reversal of
positive density ones.
Since $\vert {\delta\rho/\rho}\vert$ never exceeds unity within a void,
the linear approximation might be expected to hold for a longer period
inside a void.
Also, as voids tend to expand at faster rates than the mean Hubble flow of the
Universe, matter within voids will have a tendency to to be swept up into two
dimensional sheets separating neighboring voids.
Both these considerations have led to the construction
of a geometrical model of large scale structure based on Voronoi tesselation,
in which voids act as centers of repulsion and matter collects in sheets,
filaments and nodes, which together make up the skeleton of the large scale
structure of the Universe (Icke $\&$ van de Weygaert 1987;
van de Weygaert 1991).

In this paper we propose to study the properties of voids using a
semi-analytic approach to model non-linear gravitational instability,
known as the adhesion approximation
(Gurbatov, Saichev $\&$ Shandarin 1985, 1989; Shandarin $\&$ Zeldovich 1989).
Since the adhesion model incorporates the Zeldovich approximation, we
expect the predictions of the latter to be also valid for the former at
early times. Most voids, according to the Zeldovich approximation,
can be successfuly associated with peaks in the primordial
gravitational potential.
However, at late times, after shell
crossing and the consequent breakdown of the Zeldovich approximation,
the inherently nonlocal nature of the adhesion model
begins to manifest itself. At such times, the location and size of a void is
determined not so much by the location and height of peaks, as
by the global structure of the primordial gravitational potential.
As a result the association of peaks in the gravitational
potential with voids, which exists in the Zeldovich approximation,
begins to progressively breakdown at late times.
The adhesion model, like the
model based on Voronoi tesselation, describes the Universe in terms of a
skeletal structure consisting of pancakes (sheets), filaments, clumps and
voids.
However, barring this resemblance, important differences
exist between the two models. Namely, voids are not present {\it abinitio}
in the adhesion model but arise out of dynamical considerations.
Furthermore, the rich dynamical structure of the adhesion model allows it to
address a
number of important issues concerning voids, such as:
Do voids evolve with time~? Are the voids seen today primary (\ie first
generation) or were they formed through the merger of an earlier generation
of voids~? What is the likelyhood that voids will
be populated by substructure such as mini-pancakes~?
What is the topology of the large scale structure of the Universe~? {\it etc.}
We shall attempt to address some of these questions plus a few more in the
present paper.

The organisation of our paper is as follows:

We briefly discuss the adhesion approximation in section 2, and compare
some of its predictions with those of the Zeldovich approximation, and
N-body simulations.
In section 3 we apply the adhesion model to obtain a distribution of void sizes
-- the void spectrum, for idealised primordial spectra
$P(k) \propto k^n, ~n = -1, 0, +1$ (Melott $\&$ Shandarin 1993),
as well as for more
realistic initial conditions  attempting
to explain the large scale structure of the Universe, such as the cold dark
matter model.
We find that the void spectrum evolves with time -- the mean void diameter
increasing as the Universe expands. We also find that a distinct correlation
exists between the mean void size and the value of the primordial gravitational
potential at the void center -- larger voids being associated with higher
regions of the primordial potential.
Our results also show that, in several instances a void can be multiply
connected indicating a nontrivial topology, with
minor pancakes (or filaments) running through a void that is
circumscribed by major Zeldovich pancakes. The void topology evolves with time,
proceeding from a sponge like topology at
early times, to a bubble like topology at late times.
Our results concerning the void topology are summarised in Section 4.
A discussion of our results is presented in section 5.

\bigskip
\vskip .5cm
\centerline{II. \phantom {.} THE ADHESION MODEL AND THE ZELDOVICH
APPROXIMATION}
\v4
The adhesion model is a logical extension of the Zeldovich
approximation, which describes the evolution of density perturbations
in a collisionless, self-gravitating medium, until the epoch of caustic
formation and shell crossing (Zeldovich 1970; for review see Shandarin
$\&$ Zeldovich 1989). We assume a spatially flat, matter dominated
Universe: $\Omega_m=\Omega =1, \Lambda =0$.

Particle trajectories in the Zeldovich approximation, are described by the
mapping
$$
\vec x(\vec q, t) = \vec q - a(t) \vec \nabla \Phi
\eqno(1)
$$
where $\vec q$ is the initial (\ie Lagrangian) comoving coordinate, and
$\vec x$  is the final
(\ie Eulerian) comoving coordinate of a particle. $a(t)$ is the scale factor
which coincides with the growing mode of the gravitational instability
in a matter dominated spatially flat Universe.
$\Phi$ is the linear velocity potential related to the
primordial Newtonian gravitational potential $\phi$ by
$$
\phi =  A^{-1} \Phi
\eqno(2)
$$
where $A = 2/(3 H^2 a^3),$ $H$ is the Hubble parameter: $H^2\propto a^{-3}$
implying $A$ is a constant.
The Poisson equation
$$
{\delta\rho\over\rho} =  A a(t) \nabla^2 \phi
\eqno(3)
$$
together with the linear growth law $ {\delta\rho/ \rho} \propto a(t),$
implies that the linearised Newtonian potential $\phi$, like its counterpart
the velocity potential $\Phi$, does not depend upon time.

The continuity equation $\rho(x) d^3x = \rho(q) d^3q$ ($\rho(x)$ is
the Eulerian density, and $\rho(q) = \bar\rho = constant$ is
the Lagrangian density), when combined with the Zeldovich approximation (1)
describes the deformation of an infinitesimal volume element
in terms of the Jacobian of the transformation from $\vec q$ to $\vec x$,
so that
$$ {dV_E\over dV_L} = J \left [{\partial \vec x\over \partial \vec q}
\right ] =
\left \vert \delta_{ij} - a(t) {\partial^2\Phi\over \partial q_i\partial q_j}
\right \vert \eqno(4a) $$
and, as a result
$$ dV_E = dV_L \left [1 - a(t)\lambda_1(q)\right ]
\left [1 - a(t)\lambda_2(q)\right ] \left [1 - a(t)\lambda_3(q)\right ]
\eqno(4b)
$$
where $dV_E$ and $dV_L$, are the volume elements in Eulerian space
(E-space) and Lagrangian space (L-space) respectively,
and $\lambda_1, \lambda_2$ and $\lambda_3$ are the eigenvalues of the
deformation tensor $\psi_{ij}:$
$$\psi_{ij}\equiv {\partial^2\Phi \over \partial q_i \partial q_k}.$$

 From (4b) we see that one (or more) of the eigenvalues ${\lambda_i}$ have to
be
positive, in order for (4) to describe a contraction
along one (or more) axis $(\ie ~~dV_E < dV_L)$. On the other hand all three
eigenvalues $\lambda_1, \lambda_2, \lambda_3$ must be negative, in order for
the volume element to expand along all three axis $(\ie ~dV_E > dV_L)$.
It is well known that in the Zeldovich approximation, the condition
$(dV_E < dV_L)$, ultimately gives rise to the formation of caustics
such as pancakes, filaments and clumps, whereas the opposite condition
$(dV_E > dV_L)$ determines voids. The validity of (4) breaks down soon after
the formation of caustics, which occur when
$a(t) = min \lbrace{ 1 /\lambda_i}\rbrace $.
Thereafter the thickness of pancakes grows unbounded in the Zeldovich
approximation, which runs counter to the findings of N-body simulations which
show that the thickness of pancakes remains considerably smaller
than both the pancake size as well as the mean distance between pancakes.
In order to stabilise the thickness of pancakes, a mock viscosity term,
which mimicks the effects of nonlinear gravity, must be incorporated into
the Zeldovich approximation.

To see how this may be done, consider the Euler equation
$$
{\partial\vec v\over\partial t} + {1\over a}(\vec v \ \nabla)\vec v + H\vec v =
- {1\over a} \nabla \phi
\eqno(5)
$$
where $$\vec v = a{d\vec x\over dt},$$ $\vec x$ being the comoving coordinate
of the particle (or fluid element).
In terms of the new comoving velocity variable
$$\vec u = {d\vec x\over da} = {\vec v
\over a \dot a},$$ (5) can be recast as (Kofman 1991)
$$
{\partial\vec u\over\partial a} + (\vec u \ \nabla)\vec u = - {3\over 2 a}
(\vec u + A \nabla\phi)
\eqno(6)
$$
The right hand side of equation (6) is reminiscent of a force term,
in the Lagrangian
approach. Setting it to zero we obtain the Zeldovich approximation.
In the adhesion approximation, we replace the right hand side of equation
(6) by a
viscosity term which mimicks the adhesive effects of nonlinear gravity.
As a result the Euler equation for the velocity field becomes
$$
{\partial\vec u\over\partial a} + (\vec u \ \nabla)\vec u =
\nu~\nabla^2\vec u
\eqno(7)
$$
which is the Burgers equation (Burgers 1974).
For potential flows equation (7) has the following exact analytical solution
$$
\vec u(\vec x, a) =
{{\int d^3q~ {(\vec x -\vec q)\over a(t)} ~ \exp \left ({1\over 2\nu}
G(\vec q, \vec x, a) \right )}\over {\int d^3q ~\exp \left ({1\over 2\nu}
G(\vec q, \vec x, a) \right )}}
\eqno(8)
$$
where
$$G(\vec q, \vec x, a) = \Phi(\vec q) - {(\vec x -\vec q)^2\over 2 a(t)},$$
$\Phi(\vec q)$ is the potential of the linear velocity field introduced
earlier: $\vec u(\vec x, a = 0) = - \vec \nabla \Phi$.
For small values of the viscosity parameter $\nu \rightarrow 0$, the
integrals in equation (8) can be evaluated using the method of steepest
descents. This leads to
the following condition being imposed on the gradient
of the gravitational potential:
$$
\nabla_{\vec q} P(\vec q;\vec x,a) =  \nabla_{\vec q} \Phi (\vec q)
= A \nabla_{\vec q}\phi(\vec q)
\eqno(9)
$$
where $$P(\vec q;\vec x,a) = {(\vec x - \vec q)^2\over 2 a(t)} + P_0.$$
Equation (9) has an elegant geometrical interpretation which can be
summarised as
follows: Whether or not a particle originally located
at $q_0$ is stuck within a pancake
can be found by descending a paraboloid $P(q;x,a)$ with radius of
curvature $2 a(t)$ and height $P_0$
onto the linear gravitational potential $\phi$,
in such a manner so as to be tangential to the potential at $q_0$
(see Fig.~1).
(The height of the paraboloid $P_0$ is a free parameter uniquely determined
by the condition that $P(\vec q, \vec x, t)$ osculates $\phi (\vec q)$
at $q_0.$ We use $\phi$ and $\Phi$ interchangeably, since $\phi \propto \Phi$
if the Universe is flat and matter dominated).
If the paraboloid so constructed touches or intersects the potential at
any other point $q \neq q_0$, then we say that the particle in
question has already entered a caustic (Fig.~1, middle and bottom panels),
otherwise it has not
(Fig.1, top panel). Thus, one can divide the Lagrangian space at any given time
into
{\it stuck} and {\it free} regions. Stuck regions correspond to particles
already in caustics, whereas free regions are still expanding via the
Zeldovich relations (1,4) and correspond to voids.
The apex of the paraboloid $\vec x,$ which osculates the potential
in two or more points (without intersecting elsewhere) describes
the location of the caustic in Eulerian space.
The mass in caustics (clumps) in Fig. 1 is given by
$m_1 = \vert q_1 - q_2\vert~\bar\rho, ~~
m_2 = \vert q_3 - q_4\vert~\bar\rho, ~~$
and $M = \vert q_1' - q_4'\vert~\bar\rho,$ $\bar\rho$ being
the mean density. (The clumps are drawn so that the clump radius
is proportional to its mass.)
The velocities of the caustics can be found from
$\vec u = {d\vec x/da} =
(\phi_{q_i} - \phi_{q_j})/ (q_i - q_j)\times \vec n$, where
the unit normal $\vec n$ points from a higher value of the
gravitational potential towards a lower value.
(For earlier implementations of the adhesion model using the geometric
approach, see Kofman, Pogosyan $\&$ Shandarin 1990, Sahni 1991,
Williams et al. 1991, Kofman et al. 1992.
For an alternative implementation of the adhesion approximation see
Nusser $\&$ Dekel 1990, and Weinberg $\&$ Gunn 1990.)

The skeleton of the large scale structure
can be derived from the division of {\it L-space} into stuck and free regions,
by noting that the border between these regions consists of particles
which are only {\it just} entering into caustics. Since the Zeldovich
approximation is valid right until a particle ends up in a caustic,
we can determine the precise location of caustics in {\it E-space} by
moving the border between stuck and free Lagrangian regions by means of the
Zeldovich approximation. We do this for different values of $\sigma (t)$
defined in units of the {\it rms} linear density contrast on the grid scale:
$$\sigma (t) = \left \langle (\delta \rho / \rho)^2 \right
\rangle^{\half} \equiv \left [a^2
\int_0^\infty{ P(k) k^{N_D-1} dk}\right ]^{\half}, \eqno (10)$$
where $N_D$ is the dimensionality of the space.
Our results are shown in Fig. 2 a-e.
In Fig.~2a, b the distribution of caustics (dots)
was obtained by moving the
border between stuck (unshaded)
and free (shaded) Lagrangian regions by means of the Zeldovich
approximation (equation (1)). The area of clumps
in Fig.~2c is drawn proportional to the clump mass obtained using the
adhesion approximation.
Comparison of the adhesion model (Fig.~2b), with the two-dimensional
N-body simulations of
Melott and Shandarin 1989 (Fig.~2d), shows very good agreement at an
epoch when the Zeldovich approximation (Fig.~2e) breaks down.

An interesting question concerns the percolation of {\it stuck} and
{\it free} regions in L-space. From Fig.~2 a-c we clearly see that for small
values of $\sigma$ ($\sigma < \sigma_1$)
the free phase percolates (Fig.~2a),
whereas for large values of $\sigma$, ($\sigma > \sigma_2$),
percolation occurs in the {\it stuck} phase (Fig.~2c). It would be interesting
to estimate
the values of $\sigma_1, \sigma_2$, and their dependence on the power spectrum
$P(k)$. It would also be interesting to determine whether
an epoch exists, when both
phases percolate (\ie whether $~ \sigma_2 < \sigma_1$).
These questions might shed some light on the issue of the large scale topology
of the Universe (\ie whether ``bubble-like" or ``sponge-like"), and will be
examined in detail in a forthcoming paper (Sahni, Sathyaprakash $\&$
Shandarin 1993).

The geometrical picture of the adhesion model, allows us to make some
general statements regarding the relative concentration of matter in
clumps, filaments, pancakes and voids, which are the structural units
constituting
the skeleton of the large scale structure of the Universe in three dimensions.
The whole process of structure formation can be viewed as a mapping of the
space at the initial time when the density is virtually homogeneous
(the Lagrangian
space) into the space at the time in question (the Eulerian space).
All structural units except voids have at least one
dimension $\varepsilon$ (thickness) which is much smaller than the other
dimensions $\varepsilon \ll L$ .
$\varepsilon$ is determined by the relaxation
processes and in the adhesion model is parametrised by the viscosity
coefficient $\nu$. The limiting case $\nu \rightarrow 0$ is accompanied
by $\varepsilon \rightarrow 0$
(for the relation between $\varepsilon$ and $\nu$ see
Gurbatov, Saichev \& Shandarin 1989)
and, as a result, clumps become points,
pancakes become sheets of infinitesimal thickness etc. In this case, at
late times, the mapping becomes formally degenerate:
clumps correspond to the mapping
of three-dimensional volumes onto zero-dimensional points ($3D \rightarrow
0D$), filaments correspond to the mapping of two-dimensional surfaces onto
one dimensional lines ($2D \rightarrow 1D$), pancakes correspond to the
mapping of one-dimensional lines onto two-dimensional surfaces
($1D \rightarrow 2D$), and voids ($0D \rightarrow 3D$).

The relative concentration of matter within these structural units can be
qualitatively characterized
by two numbers ${{\cal D}_L, {\cal D}_E },$ specifying the type of mapping
${{\cal D}_L  \rightarrow {\cal D}_E }:$ ${\cal D}_E$ is the
dimension of the object in E-space, which is formed from matter
originally spread uniformly in a ${\cal D}_L$ dimensional volume in L-space.
Of course, the mapping of a higher dimensionsional set onto
a lower dimensional set
(clumps and filaments) formally results in infinite density, however of
different types. Clumps, corresponding to the $3D \rightarrow 0D$ mapping
are much
higher concentrations of mass than filaments which result from the
$2D \rightarrow 1D$ mapping. In addition, filaments originating from a $2D$ set
in $3D$ space have infinitesimal mass which means that they eventually vanish
in real systems. Voids  and pancakes result from the
$0D \rightarrow 3D$ and $1D \rightarrow 2D$ mappings respectively and
therefore tend to have both infinitesimal densities and masses at later
times.

This discussion may be useful in providing a guideline for a qualitative
understanding of
the evolution of the structural units with time as well as for comparing
$3D$ and $2D$ systems. In two-dimensional space clumps correspond to
the $2D \rightarrow 0D$ mapping, filaments to the $1D \rightarrow 1D$
mapping and voids to the $0D \rightarrow 2D$ mapping. Therefore one may
expect filaments in two-dimensional simulations
($1D \rightarrow 1D$) to appear
relatively more conspicuous when compared to pancakes
($1D \rightarrow 2D$) in
three-dimensional simulations. On the other hand filaments in
three-dimensional simulations ($2D \rightarrow 1D$)
must look more noticeable than their counterparts in two dimensions.
We believe that this type of reasoning is worth keeping in mind when
extrapolating (even qualitatively) conclusions from $2D$ to $3D$.

\vskip .5cm
\centerline{III. \phantom {.} THE VOID SPECTRUM AND ITS EVOLUTION}
\vskip .4cm
It is worth contrasting the dynamical picture of voids which emerges in
the adhesion model, with the essentially static picture that follows from
the Zeldovich approximation.
In the Zeldovich approximation, voids are associated with Lagrangian regions
in which all the eigenvalues of the deformation tensor are negative.
This implies essentially, that there is a one to one correspondence between
voids and peaks in the primordial gravitational potential.
This is also true at early times in the adhesion model when the paraboloid
is very narrow and can only feel small localised regions of $\phi$
(see Fig.~1, top panel).
The adhesion model at such times gives essentially the
same results as the Zeldovich approximation.
At late times however, the radius of curvature of the paraboloid can
become comparable to that of the initial gravitational potential,
with the result that the paraboloid is now able to {\it survey} larger
regions of the potential. It is at this stage that the essentially
nonlocal nature of the adhesion model emerges.
As a result, a free region in L-space
(a void) can become {\it stuck} at late times, even if it
lies in the vicinity of a peak in the gravitational potential (see Fig.~1,
bottom panel).
As a result, no strictly one to one relationship exists between voids
and peaks in the gravitational potential, in the adhesion model.
In addition, voids in L-space tend to shrink in the adhesion approximation,
whereas in the Zeldovich approximation, their volume in L-space remains
constant. (By contrast the total volume of voids in E-space remains
constant in the adhesion model.)
The distinction between voids in the Zeldovich approximation and in the
adhesion model in two-dimensions, is brought out clearly in Fig.~3
in which we superimpose voids in the Zeldovich approximation with those
obtained using the adhesion model.
 From Fig.~3 we see that virtually all
of the voids in the adhesion model, constitute a subclass of voids in the
Zeldovich approximation, although this is strictly true only at very
late times, when most of the matter is already in pancakes, so that the volume
(equivalently, area in 2D)
occupied by voids in L-space, is very small.
The terms {\it early} and {\it late} can be quantified in relation to the
{\it rms} linear density contrast on the grid scale $\sigma (t)$ defined
in equation (10),
which evolves in proportion to the
expansion factor of the Universe. (The adhesion picture
plotted in Fig.~3, corresponds to $\sigma = 8$.)

The geometrical ansatz which allows us to divide L-space into stuck and free
regions, also provides us with a very elegant way to determine both the
individual void size, as well as the entire void spectrum. As discussed in the
previous paragraph, a void in L-space is a region, matter from which
has not yet fallen into a caustic. Since the Zeldovich approximation is
still valid in such regions, one can use the volume deformation formula
given by equation (4b),
to determine the Euler volume occupied by a unit cubical element in L-space.
Summing over all such elementary volume elements within a given Lagrangian void
gives the volume of the void in E-space:
$$ V_E =
\sum dV_E =
\sum_{i=1}^N
dV_L~
\left [ 1 - a(t)\lambda_1(q_i)\right ]
\left [ 1 - a(t)\lambda_2(q_i)\right ]
\left [ 1 - a(t)\lambda_3(q_i)\right]
\eqno(11)$$
where $N$ denotes the number of elementary {\it free} volume elements in a
given L-space void.

(Note that voids in L-space are defined on a discrete grid, whereas those in
E-space are {\it not}. Consequently $V_L = N dV_L = N ({\lambda_N/2})^3$
gives the volume of the void in L-space, where
$\lambda_N$ is the Nyquist wavelength.)

This method, which is a useful synthesis of both the adhesion as well as the
Zeldovich approximations, allows us to determine the precise volume of a
void without resorting to any simplifying assumptions regarding its shape,
such as sphericity, {\it etc.}, though it neglects the volume occupied
by clumps, filaments and pancakes.
We apply this ansatz to determine the spectrum of void sizes, for a Cold Dark
Matter model with $\Omega = 1,$ $H_0 = 50$ km/sec/Mpc,
and amplitude normalised by the COBE observed CMBR anisotropy
(The COBE - normalised amplitude gives $(\Delta M / M) \simeq 1.1$
for the {\it r.m.s.} CDM mass fluctuation on scales of $8 h^{-1}$ Mpc
(Smoot et al. 1992, Efstathiou, Bond $\&$
White 1992). The corresponding value of
$\sigma (t)$ is 9.)\footnote{$^2$} {
In normalising the CDM spectrum, we have ignored the gravity wave contribution
to $(\Delta T/T)$, which in some cases can be significant
(Davis et al. 1992, Liddle \&Lyth 1992, Lidsey \& Coles 1992,
Lucchin et al. 1992, Sahni \& Souradeep 1992, Salopek 1992,
Souradeep \& Sahni 1992).}.

The results of our simulations, which average three realisations of the
CDM spectrum using $128^3$ particles, in a
128 Mpc box with periodic boundary conditions, are
shown in Fig.~4  for different cosmic
epochs corresponding to: (a) the past  $a = 0.5\ a_0,$ $z = 1$ (top panel);
(b) the present  $a = a_0,$ $z = 0$ (middle panel); and (c) the future ($!$)
$a = 1.5 \ a_0$ (bottom panel);
($a$ -- is the expansion factor, and $z$ the redshift).
We plot both the number fraction of voids (right panels)
having a given diameter: $n(D)/ N,$
$ N = \sum_i n(D_i)$, as well as the associated void volume fraction
(left panels):
$v(D)/ V = n(D)\times V_E(D)/V,$
$V = \sum_i V_E(D_i)$.
(The void diameter in our simulations is defined so that
$4\pi ({D/ 2})^3/3 = V_E(D)$,
where $V_E$ is the void volume in
Eulerian space, described by equation (11).)
The void spectrum in the middle panel of Fig.~4 has
been plotted for $\sim 3400$ voids,
 -- the mean number of voids at the present epoch in our simulation.
 We find that the volume spectrum of voids shows a fairly uniform spread for
 voids having diameters in the range $10 \le D \le 30$ Mpc where $D$ is the
 void diameter.
 Approximately $65\%$ of the total volume in voids, is contained in voids
 lying within this range.
 By contrast, the number spectrum of void sizes, has a well defined peak
 on scales of $\sim 6$ Mpc, indicating that the most abundant voids in our
 simulation ($\sim 50\%$) are of this size.
 However, the net contribution from
 such small size voids, to the overall void volume is a tiny $\sim 10\%$.

 From Fig.~4
we see a distinct evolution of the void spectrum with
cosmological epoch, the average size of a void growing roughly with time as
$ \bar D(z) = {\bar D_0 / \sqrt {1+z}},$ where
$\bar D_0 \simeq 10.5$ Mpc, is the mean diameter of a void today and $z$ is the
cosmological redshift.
\footnote{$^3$} {The mean void diameter is evaluated using
 $4\pi (\bar D/2)^3/3 = \bar V$,  where
 $\bar V = N^{-1} \sum_{i=1}^N V_i$
 is the mean void volume. Taking an alternate
 definition for $\bar D:$ $\bar D = N^{-1} \sum_{i=1}^N D_i$ results in a
 somewhat smaller value $\bar D \simeq 7$ Mpc. \hfill \break}
(The modal diameter of a void in our simulations
turns out be slightly larger: $D_{mode} \simeq 17$ Mpc.
The maximum void size in our simulations is $\simeq 60$ Mpc.)

The void spectrum obtained by us using the adhesion model, may be compared with
the void spectrum reconstructed by KF  from a
preliminary catalogue of 129 voids .
(The voids in our simulations are empty regions, the density of matter
in a void being less than $1\%$ of the mean density of matter in the
Universe.
Likewise the voids in the KF sample are also assumed to be
empty.)

KF obtain $D_{mode} \simeq 8 - 11 h^{-1} \ {\rm Mpc,}$
$h = {H/100}$, for the typical (viz modal) diameter of a
void in their catalogue.
Applying the same criteria as KF, and considering only
voids having diameters greater than $5 h^{-1}$ Mpc in our sample, we find
the mean, median and modal void sizes in our simulations to be
$D_{mean} = 19.1 \pm 0.3$ Mpc,
$D_{median} = 26.4 \pm 0.8$ Mpc, and
$D_{mode} = 25 \pm 5$ Mpc
respectively, at the present epoch.\footnote{$^4$} {
The mean, median and mode are defined with respect to the void volume
fraction. Voids having the modal void size therefore provide the largest
contribution
to the overall void volume spectrum. Evaluating statistical averages
for the void number spectrum, gives much smaller values both
for the median as well as the modal void size
due to the abundance of small voids in the CDM simulation.}
As a result we find that the modal diameter of a
void in our sample is just about the same as that in the KF sample.
(However, the mean diameter of a void in our simulations appears to be
smaller than that implied in the KF survey.)
 The good agreement shown between the observed modal
 void size and that predicted
 by the CDM model must however be treated with some caution,
since (a) the modal void size in our CDM simulations is
subject to fairly large uncertainties (in fact, we do not have a well defined
modal value at all) and
(b) the methods we employ to determine
the void size and those employed by KF are somewhat
different.
Namely, KF
work with an uncontrolled data set and may therefore be systematically
overestimating void sizes. KF further make the assumption that voids are
essentially compact so that two neighboring voids do not percolate,
they then determine the void size by inscribing the given void with a cube
-- the edge length
of the cube providing a measure of the void size.
While this method has the advantage of providing a well defined
implementational algorithm to determine void sizes, it suffers from the
drawbacks that: (i) voids are assumed to be approximately ellipsoidal
with comparable axes, and
(ii) because voids are assumed to be isolated, the possible sponge-like
topology of voids is ignored.
Neither of these assumptions is made in the adhesion model, in which the void
volume is calculated for voids which can have virtually any shape
and topology --  a point that we shall elaborate on in the next section.
Consequently, in order to rule out (or confirm) the predictions made by the
CDM model, we must apply the same void finding algorithm
both to the data as well as to the
simulations. We propose to do this in a follow up paper.

An important related issue which we have been unable to adequately settle in
the present paper concerns the distribution of void sizes in a realistic (\ie
spatially unbounded)
CDM Universe. Simulations performed by us
on $64^3 \ {\rm Mpc}^3$ and $128^3 \ {\rm Mpc}^3$
sample volumes of a CDM Universe
show that, although larger voids are present in the $128^3$ sample relative to
the $64^3$ one, the mean void size does not change much. We get a value of
$D_{mean} \simeq$
10 Mpc (18 Mpc)  in a $64^3$ sample and $D_{mean} \simeq$ 10.5 Mpc (19 Mpc)
in a $128^3$ sample.
(The numbers quoted in parentheses are those obtained after discarding
all voids
of size less than 10 Mpc as in KF.)
This could indicate that the mean diameter of voids converges rapidly in an
$\Omega = 1$ CDM
Universe so that the results which we quote for $D_{mean}$ from our
$128^3$ simulations could be representative of the entire
CDM Universe. However, this issue can be completely resolved only after
larger simulations have been performed.
(Incidentally, the maximum void size increases from
$D_{\rm max} \simeq 40\pm 8$ Mpc in a $64^3$ simulation to
$D_{\rm max} \simeq 57\pm 6$ Mpc in a $128^3$ simulation.)
We would also like to mention in this connection, that the median (evaluated
with respect to the void number spectrum) is usually a
considerably more stable statistical indicator of void sizes than the mean.
The reason for this is that the median is much less sensitive
to the presence of a few very large voids in a sample than the mean,
and consequently its value
does not grow as rapidly as the latter as we increase the sample size of
our model Universe.  (See also Kauffmann \& Melott 1992).

Fig.~1 shows that one might expect to see a correlation between the sizes of
voids, and the height of the linear gravitational potential,
especially at late times.
This feeling is borne out by the results of our
simulations for the CDM model (see Fig.~5), which show that
a distinct correlation exists between the size of a void, and the value of the
gravitational potential evaluated at its center.
The value of the potential averaged over all the void centers at the present
epoch is
about 0.75 times the {\it rms} value of the potential, indicating that
voids preferentially form in those regions where
the gravitational potential is large.
In order to demonstrate the general nature
of this relationship between void size
and the value of the gravitational potential $\phi$,
we have also analyzed simulations for power law primordial spectra:
$P(k) = \langle\dk2 \rangle \propto k^n$,
where $\vert\delta_k\vert$ is the spatial Fourier transform of the density
contrast (see Melott \& Shandarin 1993).
It is assumed that the phases of the Fourier components
are randomly distributed, so that the gravitational potential
$\phi(x)$ has the statistical properties of
a Gaussian random field. A cutoff was introduced into the spectrum
by requiring that $P(k) = 0$ for $k > k_c$. (Physical processes which
can give rise to such a cutoff include the free streaming of
weakly interacting massive particles (WIMP's) such as massive neutrino's,
and the Silk damping of inhomogeneities in the photon-baryon plasma.)
Our simulations were run on a $128^3$ box
for the following three models: $n = -1, 0, +1$, and
$k_c=16\times k_f$. ($k_f$ is the wavenumber
of the fundamental mode: $k_f = {2\pi/L}$ where $L$ is the length of the
side of the simulation cube.) Three simulations were performed for each
of the three spectra amounting to 9 simulations in all.

Our results are shown in Fig.~6 for different values of $\sigma (t).$
The upper limit in the integral in equation (10) is
effectively replaced by $k_N$ -- the Nyquist frequency in the simulation.
$\sigma (t)$ can be related to $k_{NL}$ -- the wavenumber of a mode
going nonlinear at the given epoch:
$$
\left \langle \left ({\delta\rho/\rho} \right )^2 \right \rangle
\equiv {a^2\over 2 \pi^2} \int_0^{k_{NL}} P(k) k^2 dk = 1
\eqno(12)
$$
so that
$$ \sigma(t) =
\left ( {{ \int_0^{k_N} {P(k) k^2 dk}\over \int_0^{k_{NL}} {P(k) k^2 dk}}}
\right )^{1/2}.
\eqno(13)
$$
We see that for truncated power law spectra ($n > -3$)
$$
\sigma(t) = \left ({k_c\over k_{NL}}\right )^{{n+3}\over 2},
\eqno(14)
$$
consequently, $k_{NL}$ decreases
with epoch, as successively larger scales enter the nonlinear regime.
We do not evolve our simulations beyond the epoch when the largest
scale to go nonlinear equals the box size.

 From Fig.~6 we find that as in the case of the CDM model,
a marked correlation
exists between the diameter of a
void and the value of the gravitational potential at its centre at late times.
We also find that the mean value of the potential evaluated at void centers
$$\langle \phi \rangle = {1\over N} \sum_{i=1}^N \phi_i, $$
$N$ being the number of voids in a simulation, increases
with time, being larger at a given instant of time, for steeper spectra
(see Fig.~7).
${\langle \phi \rangle/\phi_{rms}} \propto \sqrt{\sigma(t)} $
provides a reasonably good
approximation, especially at late times, to the rate of growth
of $\langle \phi\rangle $ for the power law spectra considered by us.

 From figures 5, 6 we find that the correlation between diameter and
potential is more striking for spectra with more short range power (such
as the truncated $n = 1$ spectrum).
This may be because a potential having significant long range power
(such as CDM) has features on it (small scale {\it wiggles} superimposed on
long range {\it mountains} and {\it valleys}) which will
influence the sites of void formation. As a result a small wiggle in
the potential will generally give rise to a small void even if the wiggle
occurred at a large value of ${\langle \phi \rangle/\phi_{rms}}$.
(The dip in ${\langle \phi \rangle/\phi_{rms}}$ occuring for void
dimaters $10 - 20$ Mpc. in figures 5, 6, which is more
pronounced for spectra with more large scale power, is a
distinctive feature of this effect.)
This modulation in the void spectrum caused by low frequency waves in
the potential, will lead to a more noisy relation between
the void diamter and ${\langle \phi \rangle/\phi_{rms}}$ for spectra
with significant large scale power, as is indeed demonstrated in figures 5,6.

The correlation of void sizes with the ``height" of the gravitational potential
seen in Fig.~5 and 6, raises the interesting possibility of
reconstructing the primordial form of $\phi$ from the observed form of the
void spectrum. The primordial form of the gravitational potential is
determined both by physical processes occuring in the very early Universe
(such as Inflation), as well as by the nature of dark matter, hence
some indications as to its form would be of immense value
(see also Kofman \& Shandarin 1988).

The evolution of the void spectrum and the transfer of power from
smaller to larger voids at late times which was shown to occur for
the CDM model
is demonstrated with great clarity in Fig.~8 for power law spectra.
(Note particularly the evolution of the void number histogram shown
in Fig.~8 b-d and the corresponding evolution of the mean
and maximum void sizes in Fig.~9a.)
We find that for large voids $V > V_{mean}$
the number spectrum of voids in Fig. 8b - 8d is well fitted by the
exponentially decaying function $\exp (-V/V_*)$ where $V_* \propto
V_{mean}$ .
A similar result was also obtained by Kofman et al. (1992) for the two
dimensional case.

 From Fig.~9a we find that the mean void diameter grows very nearly as
$D_{mean} \propto \sqrt{a(t)}$ at late times.
(The complementary picture demonstrating the evolution
of the total number of voids in our simulations is shown in Fig.~9b.)
For the truncated scale invariant spectrum $n = 1$, this result agrees
with an asymptotic analysis of the growth of cellular structure in the
adhesion model made by Gurbatov et al. (1985, 1989), who found that
$D_{mean} \propto \sqrt{a(t)}$ if $n \ge 1$, and $D_{mean} \propto
a(t)^{2/(n+N_D)}$ ($N_D$ being the dimensionality of space) for $- 1 < n <
1$. The last result suggests that the mean void size grows faster for
spectra with more large scale power, and is seemingly at odds with our
analysis which indicates that $D_{mean} \propto \sqrt{a(t)}$ for
$-1 \le n \le 1$. This discrepency might arise because of the following
two reasons: a) In their analysis Gurbatov et al. implicitly assume
that voids are associated with peaks in the gravitational potential.
We find on the other hand, that this assumption is not always satisfied,
since voids can sometimes be associated with other features in the
potential such as ridges etc. (this is especially true in two and three
dimensions.) As a result the analysis of Gurbatov et al. is strictly
valid at asymptotically late times. b) Our simulations are carried
out in a finite sized box which might inhibit the late time growth of
voids particularly for fluctuation spectra with long range power.
A similar effect was earlier discussed by Kofman et al. (1992)
in the context of the two dimensional adhesion model.
Indeed, simulations conducted by us on spectra having no small scale cutoff
show that $D_{mean}$ grows proportionally to $a(t)^{2/(n+N_D)}$
for $n = 0, -1$ as expected.

An interesting feature which emerges from our simulations
with a cutoff in the initial spectrum is that the number
of voids
peaks at a given value of the expansion factor $\sigma(t) = \sigma_*$,
and then declines steadily (see Fig.~9b).
The value of $\sigma_*$ is relatively insensitive to
the form of the spectrum for spectra with $k_c \gg k_f$, and can be
described by the analytic formula:
$$
\sigma_* = \sigma(k_{NL}^{-1} = R_*)
\eqno(15)
$$
where $R_* = \sqrt 3\sigma_1/ \sigma_2.$ $\sigma_j$ are the
moments of the distribution of $\phi$ (Bardeen et al. 1986, Kofman 1989):
$$
\sigma_j^2 = {1\over 2\pi^2} \int dk k^2 k^{2j} |\phi_k|^2.
\eqno(16)
$$
The scale $R_*$ characterises the mean distance between
peaks of the gravitational potential.
For the truncated power law spectra considered by us:
$k_c\times R_* \simeq
\sqrt {6},~ 3, {\rm and} ~4,$ for $n = 1$, 0, and -1,
respectively. For spectra having no built-in scale, the value of $R_*$
would be determined by the small scale cutoff introduced by the
Nyquist frequency in a simulation.)

Fig.~10 shows the fraction of matter in caustics (stuck matter)
$f(\sigma)$
(dotted curves), and the related underdensity of matter in voids
$1 - f(\sigma)$ (solid curves)
as a function of epoch, for the same set of simulations.
Our simulations show that the rate of infall of matter into caustics
is virtually insensitive to the spectral index in the range $-1 \le n \le +1$.
We also find that by
$\sigma(t) = \sigma_*$, virtually all of the matter is already in caustics,
so that voids are
relatively empty by this epoch (see also Pogosyan 1989, Sahni 1991).
(The fraction of matter in caustics can be described to within $2\%$
accuracy by the fitting function
$f(\sigma) = \sigma^{3.6} (3 + \sigma^{3.6})^{-1}$ for $\sigma \ge 1$.)

Fig.~9b lends support to the viewpoint that the growth of
large scale structure in the
Universe is characterised by two complementary regimes.
The first regime describes the
formation of pancakes and the establishment of cellular structure
($\sigma(t) \le \sigma_*$
during this period). However, the emergence of cellular structure in the
Universe,
is only an intermediate asymptote, and is succeeded by a second epoch
during which large mass concentrations (knots and filaments) attract
one another leading to the disruption of
cellular structure at late times.
As a result, whereas gravitational instability leads to pancaking
during $\sigma(t) \le \sigma_*$,
hierarchical clustering characterises
the later period $\sigma(t) > \sigma_*$.
This interpretation is confirmed by a comparison of the
results of 2D N-body simulations (Fig.~11 a-c) (Melott \& Shandarin 1989;
Beacom et al. 1991) with those
of the adhesion model (Fig.~11 d-f).
 From Fig.~11 it is clear that the formation of cellular
structure is complete by $\sigma = \sigma_* \simeq 4 $ and that from then
onwards
clustering proceeds hierarchically.
In fact, from Fig.~11 a-c and d-f we find that the number of voids
first increases from one interconnecting void at $\sigma=1,$ to a
maximum of eight voids at $\sigma=4$ and then decreases to five
voids at $\sigma=16.$
(We would like to add that despite the fact that gravitational clustering
proceeds hierarchically during the second epoch ($\sigma(t) > \sigma_*)$,
we can still
find the emergence of large correlated structures such as
{\it secondary pancakes}, during this period, as has recently been
demonstrated by
Kofman et al. 1992.)
This view of the formation and evolution of large scale structure is
borne out by a study of the evolution of the void spectrum shown in
Fig. 4 and Fig.~8a-d.
 From these figures we find that there are a greater number of large voids
at early and late times.
This clearly demonstrates that the evolution of voids proceeds in two distinct
phases. During the first phase, $(\sigma(t) \le \sigma_*)$ large voids fragment
into smaller voids. During the second phase, $(\sigma(t) > \sigma_*)$, the
reverse process takes place as voids begin to coalesce into larger units.
The mean and maximum void sizes shown as functions of
$\sigma(t)$ in Fig.~9a, provide further support to this point of view.
Strictly speaking the above model assumes a cutoff in the initial spectrum
as in our simulations. However, if the initial spectrum falls down as
$k^{-3}$, as in the CDM model, then the evolution of the large scale structure
also follows the above discription.
The first caustics in this case are associated with the free-streaming
distance for dark matter particles -- in the case of a real
(spatially unbounded) Universe, or with the artificial, Nyquist
frequency cutoff -- for a numerical simulation.

Finally, we would like to comment on the fact that
the correlation of the height of the
primordial potential with the void size appears to be more noticeable at
late times when $\sigma > \sigma_*$ (see Fig.~5 and 6).
The reason again lies in the fact that the
formation of cellular structure is complete only by $\sigma \simeq \sigma_* $,
where
$\sigma_* \simeq 5$ for the power law spectra discussed in the
previous paragraph. Therefore at earlier times when $\sigma \le \sigma_*$ large
voids are still in a state of {\it becoming}, since what used to be a large
void at
an early epoch gets fragmented into several smaller voids at later times and
vice-versa.
As a result, at epochs prior to $\sigma \simeq \sigma_*$,
no definite relation as yet exists
between the sizes of large but transient voids, and the
height of the gravitational potential within them.
At very late times the gravitational potential at the centres of voids
is very nearly  uniform irrespective of their size.
(We should also note that since there are very few large voids in a given
simulation, their distribution is plagued by small number statistics
as evident from Fig.~5 and 6.)

\vskip 0.5cm
\centerline{IV. \phantom {.} SUBSTRUCTURE WITHIN VOIDS}
\vskip 0.4cm
A central result of our study of voids is that voids can be populated
by substructures such as minipancakes and filaments which run through a void.
Tis result emerges from a study of the topology of voids in Lagrangian space.
The topology of a compact manifold, can be characterised by
its genus measure $g$ which is related to the integrated Gaussian curvature
$K$ of the manifold, by the Gauss-Bonnet formula
$$
\oint~K~dA = 4\pi~(1 - g)
\eqno(17a)
$$
(For a sphere $K = r^{-2}, A = 4\pi r^2$, and therefore $g = 0.$)
The genus is related to the Euler characteristic of a manifold $\kappa$,
by the simple relation $ g = 1 - {\kappa/ 2}$.
At an intuitive level, we might say that the genus characteristic
provides a measure of the number of holes in a given manifold.
Thus, a sphere has genus = 0,
a torus $genus=1,$ a pretzel $genus = 2,$ and so on.
In our present discussion we shall be interested in evaluating the topology
of voids defined on a discrete grid in Lagrangian space. To do this we shall
find it convenient to work with the discrete analog of the Gauss-Bonnet
formula:
$$
\sum_i D_i = 4\pi (1 - g).
\eqno(17b)
$$
Here $D_i$ is the angle deficit defined at each vertex of the polyhedral void.
Equation (17b) can be derived from equation (17a) if we note that all
of the curvature
of a discrete surface is localised in each of its $N$ vertices so that
$$ K(x) \simeq \sum_i D_i \delta (x - x_i)$$
and
$$\oint {K d^2x} \simeq \oint {\sum_i D_i \delta(x - x_i) d^2x}
= \sum_{i = 1}^N D_i.$$
We apply equation (17b) to determine the topology of individual voids defined
in L-space using the adhesion ansatz.
(We would like to stress that, throughout the present discussion, the topology
that is referred to will be that of {\it individual} voids, as opposed to the
topology characterising large scale structure as a whole.
See Melott 1990 for a comprehensive
review of the subject of the topology of large scale structure.)

One of the most important results to emerge from the present analysis of voids
using the adhesion model, is that voids in L-space can have a nontrivial
topology.
(We would like to point out in this context
that the 2 dimensional L-space voids shown
in Fig.~2c and Fig.~11 e,f, are all homomorphic to circles
and have therefore an
essentially trivial $S^1$ topology.
In contrast to this the voids appearing in our three dimensional simulations
sometimes show appreciable departures from a purely
spherical (\ie $S^2$) topology.)
To illustrate what we mean, we plot one of the several topologically
nontrivial void configurations that appear in our simulations in Fig.~12.
The void has the topology of a torus in L-space
\ie the {\it free region} defining the
void has a single hole (corresponding to stuck particles)
passing through it.
As mentioned in section 2 of this paper, the Euler picture of the distribution
of caustics can be constructed from the coresponding Lagrangian picture, by
moving the border between stuck and free regions by means of the Zeldovich
approximation. This essentially amounts to moving the boundary of a void in
L-space by the Zeldovich prescription (1). Since the Zeldovich approximation is
a topology preserving transformation (Shandarin $\&$ Zeldovich 1989),
it follows that a non-simply connected
void in L-space will be mapped onto a topologically non-trivial configuration
in E-space. In particular, a void having the topology of a torus
(\ie $genus = 1$),
will correspond to a void having {\it one} minor Zeldovich pancake running
through it. In general, a void in L-space having $genus = N_g$, will
correspond in E-space, to $N_g$ minor Zeldovich pancakes running through a void
bounded by major Zeldovich pancakes.
(Observationally, such voids might convey the impression of being sponge-like.)
We also feel that mini-pancakes within voids, could be young transient
features, since our simulations show that voids tend to empty out with time,
leading to a bubble like topology for voids at late times.
The dynamical mechanism leading to the trivialisation of the void topology
could be the following: As voids expand they empty out with time and
matter from within them falls into pancakes (both mini as well as major).
As a result, mini pancakes within voids evolve with time growing progressively
more massive. At a later epoch these minor pancakes graduate into major
Zeldovich pancakes and succeed in dynamically dividing a topologically
nontrivial void into two or more topologically trivial voids.
In order to estimate the evolution of the void topology with time,
we determine the mean genus characteristic of all voids $V_i$ in a given
simulation
$$
\bar g(\sigma) = {1\over N}~\sum_{i=1}^N~~ g~(V_i)
\eqno(18)
$$
$N$ being the total number of voids at a given epoch.

The results of our simulations for the CDM model, are shown in Fig.~13 and 14.
In Fig.~13 we plot the histogram of the genus per void  against void size for
three different epochs. What is plotted is $\left < g_i \right >$ --
the total genus of all voids in a certain class divided by the number of voids
in that class:
$$\left < g_i \right > = {1\over n_i} \sum_{k=1}^{n_i} g_k,$$
where $n_i$
is the number of voids in the diameter class $D_i$ and $g_k,$
$k=1,\ldots, n_i,$
are the genus values of the individual voids in that class.
This is a good estimator of genus as a function of void diameter since the
number of voids in different classes are different.
In Fig.~14 we plot the mean genus in our simulation
as a function of $\sigma (t).$ These two figures
reveal that one out of
every fourteen voids today has a nontrivial topology. By contrast,
every eighth void had a topology that was nontrivial at a redshift of unity.
Fig.~14 clearly demonstrates how
the void topology evolves and essentially trivialises with time.
A good analytical approximation to the time evolution of the mean
genus, shown in Fig.~14 for the CDM model, is provided by
$ \bar g(\sigma) \propto \sigma^{-3/2}(t).$
For comparison we also plot the genus, against void size and as a function
of $\sigma (t),$ for the power law
spectra $P(k) = \langle\dk2 \rangle \propto k^n$ for $k \le k_c$,
$P(k) = 0$ for $k > k_c$, $n = -1, 0, +1$, in Fig.~15, $\&$ 16, respectively.
We find that in this case
too, larger voids tend to have a more complicated topology than smaller
voids, and that the overall void topology tends to simplify with time.
It is interesting to note that some voids such as the
Coma void,
do indeed seem to be populated by mini-Zeldovich pancakes,
as pointed out recently by
Park et al. 1992.

 From Fig.~13 we find that the ratio of the diameter of a void in
Eulerian space to its diameter in Lagrangian space, grows very nearly as
${D_E/D_L} \propto \sigma (t)$, regardless of the size of the void.
This relation is easy to understand if one notes that at late times
equation (11) reduces to
$$
\eqalign
{V_E =& \sum_{i=1}^N dV_L~
\left [ 1 - a(t)\lambda_1(q_i)\right ]
\left [ 1 - a(t)\lambda_2(q_i)\right ]
\left [ 1 - a(t)\lambda_3(q_i)\right]\cr
{\mathrel{\mathop{=}_{a \gg 1 } } }&
- N dV_L~ a^3(t)~ \sum_{i=1}^N {\lambda_1(q_i)\lambda_2(q_i)\lambda_3(q_i)
\over N}}
\eqno(19)
$$
so that,
$$\left ({V_E\over V_L} \right )^{1\over 3}
\equiv {D_E\over D_L} =
\sigma (t) \left \langle I_3 \right \rangle^{1\over 3},$$
where
$$ \langle I_3\rangle =
-{1\over N}\sum_{i=1}^N~\lambda_1(q_i)\lambda_2(q_i)\lambda_3 (q_i),$$
is the mean value of the invariant $I_3$ in the given L-space void.
$N$ denotes the number of {\it free} unit volume elements in a
given L-space void, and
$V_L = N\times dV_L$  is the Lagrangian volume of the void
($\sigma(t) \propto a(t)$).
 From Fig.~13 it follows that small voids in L-space remain small in
E-space and vice versa. The ratio ${V_L/V_E} = ({D_L/ D_E})^3$ is
a measure of the underdensity of voids. From this figure we find that
$$\left \langle {V_L\over V_E}\right \rangle_{z = 0}
\simeq {1\over 8}\left \langle {V_L\over V_E}\right \rangle_{z = 1},$$
indicating that voids today are $\sim 8$ times more empty than
they were  at redshifts of unity. (Note that $\sigma \simeq 9$ today, for
the COBE-normalised CDM model.)

\vskip 0.5cm
\centerline{V. DISCUSSION}
\vskip 0.4cm
We have applied the adhesion model to determine the spectrum of void sizes,
both for the COBE-normalised CDM model, as well as for models with power-law
spectra $P(k) \propto k^n$, $n=-1, 0, +1.$
The model neglects the volume occupied by clusters and superclusters, however
it does not make any a priori assumptions on the geometry and topology of
the voids. The advantage of the adhesion model is that it includes both
dynamical as well as statistical aspects of void formation.
We find that most of the
characteristics of voids, such as the mean and maximum void size
in a simulation, evolve with time. For the CDM model we find that
$\bar D = {\bar D_0/ \sqrt {1+z}},$ where $\bar D_0 \simeq 10.5$ Mpc, is
the mean void size today.
We also find that the total number of voids in a simulation, varies with
cosmic epoch, increasing to a maximum value at
$\sigma_* = \sigma(k_{NL}^{-1} = R_*)$, and then decreasing
monotonically. We feel that the expansion scale $\sigma_*$
separates two distinct
phases in gravitational instability. During the first phase ($\sigma \le
\sigma_*$), gravitational instability proceeds via the formation of
pancakes, and the subsequent establishment of cellular structure, which is
complete by $\sigma(t) = \sigma_*$.
However, pancake formation happens to be an intermediate asymptote and
during the second phase ($\sigma(t) > \sigma_*$)
gravitational instability proceeds hierarchically, leading to the eventual
disruption of cellular structure.

Our results show that the sizes of voids are stongly correlated with the
height of the primordial gravitational potential at void centers,
larger voids
forming in regions where the gravitational potential is higher.
This result indicates that knowing the observed form of the void spectrum
it might be possible to, at least partially, reconstruct the primordial form
of the potential.
Assuming the Harrison-Zeldovich initial spectrum one can easily verify
that the gravitational potential is not a homogeneous random function if
the spectrum is extrapolated to $k=0$. It means that the largest void
is probably determined by the size of a sample.

One of the most intriguing results of the present
analysis is that voids can have a topology which is nontrivial.
We find that the topology of a void depends upon its size, with larger voids
more likely to have a nontrivial topology than smaller voids.
We also find that the topology of voids, like the void spectrum evolves
with time, with voids becoming progressively emptier at later times.
Our results for the CDM model show that one out of every 14 voids is likely
to have some substructure, such as a filament or a pancake passing through it.
These results are supported by
recent observations indicating the presence of galaxies
(Westrop et al. 1992) and mini-pancakes (Park et al. 1992)
within at least
two voids.
It is likely that with the advent of deeper and more complete redshift
surveys the issue of void size and void substructure will take on a deeper
significance, as a detailed picture of the texture of voids in the Universe
emerges. Bearing this in mind, we believe that the topological indicator
of void substructure developed by us in this paper, can emerge as a key
statistical indicator for the study of voids in the Universe.

If both the shape and the amplitude of the initial
perturbation spectrum are independantly known (for instance from the
angular fluctuations of the cosmic microwave background radiation), then
void statistics in the adhesion approximation
is determined solely by the growth factor of the linear
density contrast $D(z, \Omega)$ (which determines the curvature of the
osculating paraboloid). In such a case the void spectrum can potentially be
used to determine the value of $\Omega$.

In a companion paper we shall apply the void spectrum and topology measuring
ansatz, developed in the present paper, to compare and contrast several
cosmological scenario's, including ``tilted" cold dark matter models,
cold dark matter models with a cosmological constant, and models
containing a mixture of cold + hot dark matter (Sahni, Sathyaprakash
 $\&$ Shandarin 1993).

Finally we would like to mention that the consistency of the results obtained
in this
paper was checked using the 3D adhesion code independently
developed by Dima Pogosyan.
\vskip .5cm
\centerline {\bf Acknowledgements}
\vskip .4cm
We benefitted from discussions with Dick Bond, Hugh Couchman, Bala Iyer,
Chanda Jog,
Nick Kaiser, Henry Kandrup, Lev Kofman, Blane Little, Adrian L. Melott,
Jayant Narlikar and John Peacock.
We are grateful to Dima Pogosyan for sharing with us his adhesion code and for
clarifying a number of conceptual issues.
One of us (BSS) would like to thank Abdus Salam, the International
Atomic Energy and UNESCO for hospitality at the International Centre for
Theoretical Physics, Trieste where a part of this work was carried out.
BSS would also like to thank Dennis Sciama for encouragement.
One of the authors (S.F.Sh.) acknowledges NASA grant NAGW-2923, NSF grants
AST-9021414 and OSR-9255223, and the University of Kansas GRF fund
for financial support.

\vskip .5cm
\centerline {REFERENCES}
\vskip .4cm
\n
Bardeen, J.M., Bond, J.R., Kaiser, N., \& Szalay, A.S. 1986, \ap 304, 15.
\vskip .2 cm
\n
Beacom, J.F., Dominik, K.G., Melott, A.L., Perkins, S.P., \& Shandarin,
S.F. 1991, \ap {372}, 351
\vskip .2 cm
\n
Bertschinger, E. 1985, \aps, {58}, 1
\vskip .2 cm
\n
Blaes, O., Villumsen, J. V., \& Goldreich, P. 1990, \ap {361}, 331
\vskip .2 cm
\n
Blumenthal, G. R., da Costa, L. N., Goldwirth, D. S., Lecar, M., \& Piran, T.
1992, \ap {388}, 234
\vskip .2 cm
\n
Bonnor, W. B., \& Chamorro, A. 1990, \ap {361}, 21
\vskip .2 cm
\n
Broadhurst, T. J., Ellis, R. S., Koo, D. C., \& Szalay, A. S. 1990, Nature,
{343,} 726
\vskip .2 cm
\n
Burgers, J. M. 1974, The Nonlinear Diffusion Equation (Dordrecht: Reidel)
\vskip .2 cm
\n
Davis R. L., Hodges, H. M., Smoot, G. F., Steinhardt, P. J., \&
Turner, M. S. 1992, \prl {69}, 1856
\vskip .2 cm
\n
Dubinski, J., da Costa, L. N., Goldwirth, D. S., Lecar, M., \& Piran, T.
1993, to appear in ApJ
\vskip .2 cm
\n
Efstathiou, G., Bond, J. R., \& White, S. D. M. 1992, \mnras {258}, p1
\vskip .2 cm
\n
Fillmore, J. A., \& Goldreich, P. 1984, \ap {281}, 9
\vskip .2 cm
\n
Gurbatov, S. N., Saichev, A. I., \& Shandarin, S. F. 1985, Soviet Phys. Dokl.
{30}, 921
\vskip .2 cm
\n
Gurbatov, S. N., Saichev, A. I., \& Shandarin, S. F. 1989, \mnras {236}, 385
\vskip .2 cm
\n
Harrington, P. M., Melott, A. L., $\&$ Shandarin, S. F. 1993, in preparation
\vskip .2 cm
\n
Hausman, M. A., Olson, D. W., \& Roth, B. D. 1983, \ap {270}, 351
\vskip .2 cm
\n
Hoffman, G. L., Salpeter, E. E., \& Wasserman, I. 1983, \ap {268}, 527
\vskip .2 cm
\n
Icke, V. 1984, \mnras {206}, 1p
\vskip .2 cm
\n
Icke, V., \& Van de Weygaert, R. 1987, \aa {184}, 16
\vskip .2 cm
\n
Kauffmann G., \& Fairall, A. P. 1991, \mnras {248}, 313
\vskip .2 cm
\n
Kauffmann G., \& Melott, A. L. 1992, \ap {393}, 415
\vskip .2 cm
\n
Kirshner, R. P., Oemler, A. Jr., Schechter, P. L., \& Shectman, S. A.
1981, \apl {248}, L57
\vskip .2 cm
\n
Kofman, L. A. 1989, in Lecture Notes in Physics, 332,
Morphological Cosmology, ed. P. Flin, \& H. W. Duerbeck
(Berlin: Springer-Verlag) 354
\vskip .2 cm
\n
Kofman, L. A. 1991, in Primordial Nucleosynthesis and Evolution of
Early Universe, ed. K. Sato, \& J. Audouze (Dordrecht: Kluwer) 495
\vskip .2 cm
\n
Kofman, L. A.,  Pogosyan, D. Yu., \& Shandarin, S. F. 1990, \mnras {242}, 200
\vskip .2 cm
\n
Kofman, L. A.,  Pogosyan, D. Yu., Shandarin, S. F., \& Melott, A. L. 1992, \ap
{393}, 437
\vskip .2 cm
\n
Kofman, L. A., \& Shandarin, S. F. 1988, Nature, {334}, 129
\vskip .2 cm
\n
de Lapparent, V., Geller, M. J., \& Huchra, J. P. 1986, \apl {302}, L1
\vskip .2 cm
\n
Liddle, A. R., \& Lyth, D. H., 1992, \pl {B291}, 391
\vskip .2 cm
\n
Lidsey, J. E., \& Coles, P. 1992, \mnras {258}, 57p
\vskip .2 cm
\n
Lin,  C. C., Mestel, L., \& Shu, F. H. 1965, \ap {142}, 1431
\vskip .2 cm
\n
Little, B. 1992, private communication
\vskip .2 cm
\n
Lucchin, F., Mataresse, S., \& Mollerach, S. 1992, \apl {401}, L49
\vskip .2 cm
\n
Melott, A. L. 1990, Physics Reports {193}, 1
\vskip .2 cm
\n
Melott, A. L., \& Shandarin, S. F. 1989, \ap {343}, 26
\vskip .2 cm
\n
Melott, A. L., \& Shandarin, S. F. 1993, \ap {410}, 469
\vskip .2 cm
\n
Nusser, A., \& Dekel, A. 1990, \ap {362}, 14
\vskip .2 cm
\n
Park, C., Gott, J. R., Melott, A. L., \& Karachentsev, I. D. 1992, \ap {387,} 1
\vskip .2 cm
\n
Peebles, P. J. E. 1982, \ap {257}, 438
\vskip .2 cm
\n
Piran, T., Lecar, M., Goldswirth, D.S., da Costa, L.N. and Blumenthal,
G.R., 1993, \mnras, {\bf 265}, 681.
\vskip .2 cm
\n
Pogosyan, D. Yu. 1989, Tartu preprint (Estonian Acad. Sci.)
\vskip .2 cm
\n
Regos, E., \& Geller, M. J. 1991, \ap {377}, 14
\vskip .2 cm
\n
Sahni, V. 1991, in Annals of the New York Academy of Sciences, 647,
Proceedings of the Texas/ESO-CERN symposium on Relativistic Astrophysics,
Cosmology and Fundamental Physics, 749
\vskip .2 cm
\n
Sahni, V., \& Souradeep, T. 1992, to appear in
The Proceedings of the First Iberian Meeting on
Gravity, September 21 - 27, 1992, Evora, Portugal,
ed. M. Bento, O. Bertolami, \& J. Mourao (Singapore: World Scientific)
\vskip .2 cm
\n
Sahni, V., Sathyaprakash, B. S., \& Shandarin, S. F. 1993, in preparation
\vskip .2 cm
\n
Salopek, D. S. 1992, \prl 69, 3602
\vskip .2 cm
\n
Shandarin, S. F., \& Zeldovich, Ya. B. 1989, Rev. Mod. Phys., {61} 185
\vskip .2 cm
\n
Slezak, E., de Lapparent, V., \& Bijaini, A. 1993, \ap {409}, 517
\vskip .2 cm
\n
Smoot, G.F. et al. 1992, \ap {396}, L1
\vskip .2 cm
\n
Souradeep, T., \& Sahni, V. 1992, Mod. Phys. Lett., {A7}, 3541
\vskip .2 cm
\n
van de Weygaert, R., 1991, Voids and Geometry of the Large Scale
 Structure, PhD thesis
\n
van de Weygaert, R., $\&$ van Kampen, E. 1993, \mnras {263}, 481
\vskip .2 cm
\n
Vogeley, S. V., Geller, M. J., \& Huchra, J. P. 1991, \ap {382}, 44
\vskip .2 cm
\n
Weinberg, D., \& Gunn, J. 1990, \mnras {247}, 260
\vskip .2 cm
\n
Weistrop, D., Hintzen, P., Kennicutt, R. C., Liu, C., Lowenthal, J.,
Cheng, K. P., Oliverson, R., \& Woodgate, B. 1992, \ap {396}, L23
\vskip .2 cm
\n
Williams, B. G., Heavens, A. F., Peacock, J. A., \& Shandarin, S. F. 1991,
\mnras {250}, 458
\vskip .2 cm
\n
Zeldovich, Ya. B. 1970, \aa {5}, 84
\vskip .2 cm
\n
Zeldovich, Ya. B., Einasto, J., \& Shandarin, S. F. 1982, Nature, {300}, 407
\vskip .2 cm
\n
Zeldovich, Ya. B., \& Shandarin, S. F. 1982, Sov. Astron. Lett., {8}, 67
\vskip .2 cm
\bigskip
\vfill \eject
\centerline {FIGURE CAPTIONS}
\vskip .4 cm

\leftline {\bf Fig.~1}
\vskip .2 cm
The geometrical prescription of descending a paraboloid onto the gravitational
potential in order to demarcate {\it stuck} and {\it free} Lagrangian regions,
is shown in one dimension. The peaks of the potential correspond to regions
where the eigenvalue $\lambda < 0$.
The particle having Lagrangian coordinate $q_0$ is free in the uppermost
figure, and has just entered into a caustic in the middle figure.
The middle and lower figures describe the merger of clumps.
\bigskip
\leftline {\bf Fig.~2}
\vskip .2 cm
The distribution of caustics (dots) (for two-dimensional simulations)
is plotted superimposed on {\it free}
(shaded) and
{\it stuck} (unshaded) Lagrangian regions for three distinct expansion epochs:
(a) $ \sigma (t) = 1,$ (b) $\sigma (t) = 2,$ and (c) $\sigma = 8.$
(d) Results of two-dimensional N-body simulations are
shown for the epoch $\sigma (t) = 2$ (Melott \& Shandarin 1989).
(e) Particles evolved according to the Zeldovich approximation
are shown for $\sigma (t) = 2.$
\bigskip
\leftline {\bf Fig.~3}
\vskip .2 cm
Regions in Lagrangian space where both the eigenvalues of the two-dimensional
deformation tensor are negative, are shown (dotted regions).
Also shown superimposed are voids in Lagrangian space
obtained using the adhesion model, for the same spectrum,
at very late times (shaded regions).
\bigskip
\leftline {\bf Fig.~4}
\vskip .2 cm
The void spectrum for the COBE - normalised CDM model, is plotted
for three time scales: (top) the past $a = {0.5\ a_0}, z = 1$;
(center) the present  $a = a_0, z = 0$; and (bottom) the future ($!$)
$a = 1.5\ a_0$;
($a$ -- is the expansion factor, and $z$ the redshift).
In all the pictures the x-axis is an indicator of the void diameter in Mpc.
The left hand pictures illustrate the volume fraction of voids having a
given diameter ``$D$"
(in Mpc) $v(D)/V$ sometimes called the {\it void spectrum.}
($v(D)$ is the volume occupied by voids
in a bin of size $D\pm 2$ Mpc and $V$ is the total volume occupied by all
voids which, in the adhesion model, is equal to the volume of the simulation
box.) The right hand pictures show the corresponding
number fraction of voids, also plotted against the void diameter.
The error bars in these histograms correspond to {\it rms} dispersion over
three simulations. We would like to point out that the number of large
voids in a given simulation is too small to be resolved in the number
fraction histogram. Their presence is, however, seen in the void spectrum.

\bigskip
\leftline {\bf Fig.~5}
\vskip .2 cm
The value of the normalised primordial gravitational potential
(${\phi/\phi_{rms}}$)
evaluated at the centers of voids is shown plotted against
the void diameter for the present epoch for one realisation of the
COBE - normalised CDM spectrum.
\bigskip
\leftline {\bf Fig.~6}
\vskip .2 cm
The value of the normalised primordial gravitational potential
(${\phi/\phi_{rms}}$)
evaluated at the centers of voids is shown plotted against
the void diameter for power law primordial spectra
$P(k) \propto k^n,$ for $k < k_c;$ $P(k) = 0,$ for $k \ge k_c,$
where $k_c = 16\times k_f,$~$k_f = {2\pi / L_{box}}$,
is the wavenumber of the fundamental mode.
The figures (top to bottom) correspond to three different cosmic
expansion epochs: $\sigma = 4,$ $\sigma = 8,$ and $\sigma = 16.$
 From left to right the figures correspond to
$n = 1,$ (left column) $n = 0$ (middle) and $n = -1$ (right).

\bigskip
\leftline {\bf Fig.~7}
\vskip .2 cm
The growth of ${\langle\phi(t)\rangle/ \phi_{rms}}$
is shown plotted against the cosmic
expansion factor $\sigma(t)$, for two of the three power law spectra discussed
in the
previous figure: $n=1$ (solid line) and $n = 0$ (dashed line).
\noindent
\bigskip
\leftline {\bf Fig.~8}
\vskip .2 cm
(a) The void spectrum $v(D)/V$ is plotted
against the void diameter $D$, for
the truncated power law spectrum of Fig.~6, 7 with spectral index
$n=1$ and for nine different expansion epochs.
(b) The number fraction of voids $n(D)/N$ is plotted
against the void diameter $D$ for
the power law spectrum
and expansion epochs of Fig. 8a.
(c) The void spectrum $v(D)/V$ (left) and the void number fraction $n(D)/N$
(right) are plotted against the void diameter $D$ for the truncated
power law spectrum of Fig.~6, 7 with spectral index $n=0$ and for
three different expansion epochs.
(d) The same as  Fig.~8c but with spectral index $n=-1.$
\bigskip
\leftline {\bf Fig.~9}
\vskip .2 cm
(a) The time evolution of the mean (left panel) and maximum (right panel)
void diameters,
in our simulation cube, are
shown as functions of $\sigma (t)$ for the same power law primordial
spectra as in the previous
figure: $n=1$ (solid line), $n=0$ (dashed line), $n=-1$ (dotted line).
The error bars indicate the {\it rms} dispersion from three simulations.
(b) The time evolution of the total number of voids
in our simulation cube, is
shown as a function of $\sigma (t)$ for the same power law primordial
spectra as in the previous
figure: $n=1$ (solid line), $n=0$ (dashed line), $n=-1$ (dotted line).
The error bars indicate the {\it rms} dispersion from three simulations.
\vskip .2 cm
\bigskip
\leftline {\bf Fig.~10}
\vskip .2 cm
The fraction of matter in caustics (dotted curves) and
the corresponding underdensity of matter in voids (solid curves)
are shown as functions of the expansion
factor $\sigma(t)$, for the power law primordial spectra previously considered.
(The upper (lower) dotted curve and the lower (upper) solid curve correspond
to $n = -1$ ($n = 1$).)
The dispersion resulting from different simulations is so small that it
cannot be resolved in this graph. Also, the curves are virtually insensitive
to the index of the power spectrum.
\bigskip
\leftline {\bf Fig.~11}
\vskip .2 cm
Results for two-dimensional N-body simulations involving $512\times512$
particles, are shown for different instants of time corresponding to:
$(a)~~\sigma(t) = 1,$ $(b)~~\sigma(t) = 4$ and  $(c)~~\sigma(t) = 16$.
The primordial power spectrum for these simulations was assumed to be a
truncated power law: $P(k) \propto k^2$ for $k < k_c, P(k) = 0$ for
$k \ge k_c$, where $k_c = 4\times k_f$, $k_f$ being the fundamental
frequency $k_f = {2\pi /L_{box}}$. (Melott $\&$ Shandarin 1989.)
Voids in L-space (described by shaded regions)
evaluated using the adhesion
approximation are shown for initial conditions and epochs identical
to those in Fig.~11 a-c: (d) $\sigma (t) = 1,$ (e) $\sigma (t) = 4,$ and
(f) $\sigma (t) = 16.$
\bigskip
\leftline {\bf Fig.~12}
\vskip .2 cm
A topologically nontrivial void in Lagrangian space having genus = 1
is shown from our CDM simulations.
\bigskip
\leftline {\bf Fig.~13}
\vskip .2 cm
In the left hand pictures,
the mean genus of voids belonging to a given diameter
class $\langle g_i \rangle,$ is shown plotted against the void diameter
in Lagrangian space --
$D_L$, for simulations involving the CDM model,
and for the same expansion
epochs as in Figure 5 (\ie from top to bottom: ${a/a_0} = 0.5, 1, 1.5$).
The corresponding Euler diameters of voids, $D_E,$ are shown plotted
against the Lagrangian diameters $D_L$, in the right hand pictures.
The solid line corresponds to ${D_E/ D_L} =
\sigma (t) \langle  \vert\lambda_1\lambda_2\lambda_3\vert\rangle^{1\over 3}$,
which is a constant for a given
simulation.
\bigskip
\leftline {\bf Fig.~14}
\vskip .2 cm
The mean genus characteristic evaluated for the entire ensamble of voids in
the CDM model, is
shown for the expansion epochs: ${a/a_0} = 0.5, 1, 1.5$, corresponding to
the past ($z = 1$), the present ($z = 0$), and the future.
\bigskip
\leftline {\bf Fig.~15}
\vskip .2 cm
Same as Fig.~13 but for power law spectra
$P(k) \propto k^n, k < k_c; P(k) = 0, k \ge k_c,$
where $k_c = 16\times k_f,~ k_f = {2\pi / L_{box}}$,
is the wavenumber of the fundamental mode: (a)~$n=1,$ (b)~$n=0,$ and
(c)~$n=-1.$
\bigskip
\leftline {\bf Fig.~16}
\vskip .2 cm
Same as Fig.~14 but for the power law spectra of Fig.~15.

\vfill\eject
\bye